\documentclass[twoside]{LCWS11}
\usepackage[latin1]{inputenc}
\usepackage{graphics,epsfig,color}
\usepackage{epstopdf}
\usepackage{wrapfig,rotating}
\usepackage{amssymb,amsmath,array}
\usepackage{url}
\usepackage{cite}
\usepackage{subfig}
\begin{document}
\title{Interactions of hadrons in the CALICE silicon tungsten electromagnetic calorimeter} %% 
%***********************************************************************
% AUTHORS INFORMATION AREA
%***********************************************************************
\author{Roman P\"oschl$^1$ on behalf of the CALICE Collaboration
% Optional short acknowledgment: remove next line if non-needed
\thanks{This work was funded by the program {\em Quarks and Leptons} of the IN2P3 France. }
% DO NOT MODIFY THE FOLLOWING '\vspace' ARGUMENT
\vspace{.3cm}\\
% Addresses and institutions (remove "1- " in case of a single institution)
1-  Laboratoire de l'Acc\'{e}l\'{e}rateur Lin\'{e}aire\\
Centre scientifique d'Orsay, B\^atiment 200\\ 
Universit\'{e} de Paris-Sud XI, CNRS/IN2P3\\ 
F-91898 Orsay Cedex, France
%% Remove the next three lines in case of a single institution
%\vspace{.1cm}\\
%2- Second Author's Institution - Department \\
%Address of Second Author's Institution - Country\\
}
%%***********************************************************************
% END OF AUTHORS INFORMATION AREA
%***********************************************************************
\maketitle
\begin{abstract}
The CALICE collaboration develops prototypes for highly granular calorimeters for detectors at a future linear electron positron collider. The highly granular electromagnetic calorimeter prototype was tested in particle beams. We present the study of the interactions of hadrons in this prototype. 
\end{abstract}
\section{Introduction}
The next machine after the LHC will be a linear electron positron collider at the TeV scale. This machine will allow high precision measurements to extend the scientific results of the LHC. Currently the most advanced proposal is the International Linear Collider (ILC) with a centre-of-mass energy of up to 500 GeV. An alternative at higher center-of-mass energies than the ILC is the Compact Linear Collider (CLIC).

The  e$^+$/e$^-$ collisions will lead to multi-jet final states. Their reconstruction will be based on so-called particle flow algorithms (PFA)~\cite{brvi02, thomson2009particle}. The goal is to reconstruct every single particle of the final state which requires the combination of information from each sub-detector. Particularly a perfect association of the signals in the tracking systems with those in the calorimeters is required. Charged particles are measured with the tracking system, while neutral particles can only be measured in the calorimeters. The particle flow technique requires a specific detector design. A limiting factor of the particle flow, called confusion, is due to overlapping showers. To separate the particles, calorimeters need to be of high granularity. They have to be very compact since they will be embedded within the magnetic coil of the detectors.

The CALICE collaboration~\cite{calice} designs and studies electromagnetic and hadronic calorimeters for experiments at the ILC. All detectors are developed with a common approach to obtain ultra granular calorimeters optimised to particle flow algorithm. The prototypes are tested in combined beam tests in order to study several technologies and the association of electromagnetic and hadronic calorimeters. These beam tests allow for developing reconstruction algorithms and for the validation of simulations.

\section{Silicon tungsten electromagnetic calorimeter}
\begin{figure}[h]
        {\centering 
                \subfloat{\includegraphics[width=0.4\textwidth]{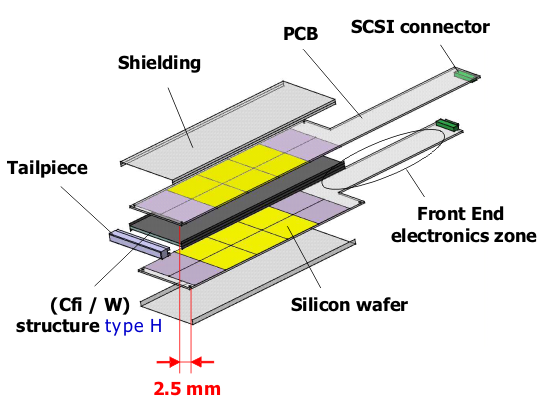}}
                \subfloat{\includegraphics[width=0.4\textwidth]{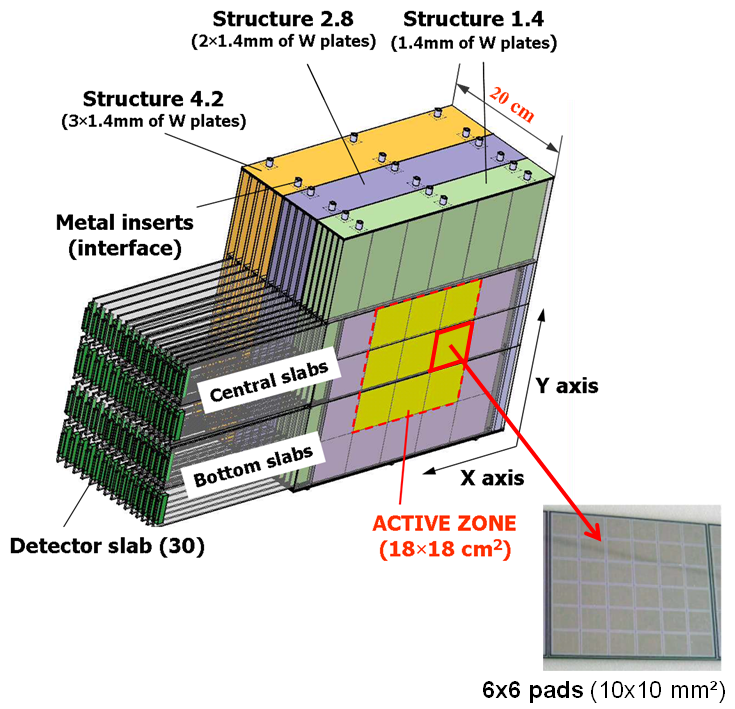}}
        \caption{\sl Pictures of the ECAL prototype. The top figure shows the structure of a single detector slab while the bottom figure shows an overview of the whole prototype.}
        \label{fig:ecalSlab}
 }
        \end{figure}

The silicon tungsten electromagnetic calorimeter (ECAL) is built following particle flow algorithm requirements and based on a sampling design. The physics prototype of the ECAL, used in this study, is composed of 30 layers of silicon as active material, alternated with tungsten as absorber material. Tungsten has a short radiation length and a small Moli\`ere radius which gives compact showers and allows for the efficient separation of close particles. The silicon allows for a thin and easily segmented readout detection system suited for high granularity. The silicon wafers are segmented in 1 $\times$ 1 cm$^2$ pads. Each wafer consists of a square of 6 $\times$ 6 pixels and each layer is a matrix of 3 $\times$ 3 wafers resulting in an active zone of 18 $\times$ 18 cm$^2$ giving a total of around 10000 readout channels. As shown on Figure~\ref{fig:ecalSlab}, the printed circuit boards (PCB) are mounted two by two in an elementary detection unit called slab. The silicon wafers are located on each side of a H-shaped tungsten supporting structure. These slabs are shielded by aluminum foils and inserted in the a carbon fiber composite mechanical structure. Three different longitudinal samplings were used, finer in the first ten layers (1.4 mm plates of tungsten), 2.8 mm plates in the ten intermediate layers, and 4.2 mm in the last ten layers~\cite{repond2008design}.

During the years 2005-2011, the calorimeter was tested with various beams: electrons, positrons, muons, pions and protons in the momentum range between 1 and 180 GeV/c. Beam campaigns were conducted at DESY, CERN and FNAL. The main goal of these beam tests was to demonstrate the principle of highly granular calorimeters. Using data collected with electron beams at energies from 6~GeV to 45~GeV, the linearity and resolution of this calorimeter to electrons were estimated~\cite{adloff2009response}. The linearity is better than 1\% while the relative energy resolution is estimated to be $\sigma_E/E = (16.53\pm0.14(\mathrm{stat})\pm0.4(\mathrm{syst}) )/\sqrt{E({\mathrm{GeV}})} \oplus (1.07\pm0.07(\mathrm{stat})\pm0.1(\mathrm{syst}))$~(\%).  To compare the test beam data with the simulation, the detector has been simulated within the Geant4-based MOKKA simulation framework~\cite{mokka}~\cite{agostinelli2003geant4}.

In addition to standard beam test analysis (calibration, noise and stability), the high granularity allow to study hadronic interactions in the detector and to validate hadronic model implementations in Geant4 simulations. As hadronic interactions are poorly understood, the high granularity of the calorimeter offers unprecedented information to study hadronic interactions. Several model combinations called physics lists are proposed in Geant4. Our studies allow for a detailed testing of these simulations. A study of interactions of pions in the physics prototype has been published in~\cite{calice2010} for energies between 8\,GeV and 80\,GeV. The work presented in this article concentrates rather on energy range 2\,GeV - 10\,GeV.

\section{Results}

The high granularity permits detailed view into hadronic showers. In order to exploit the imaging of hadronic showers, we have to develop new analysis methods like particle tracking or interaction vertex localisation in the calorimeter.

\begin{figure}[h]
        {\centering 
                \includegraphics[width=0.47\textwidth]{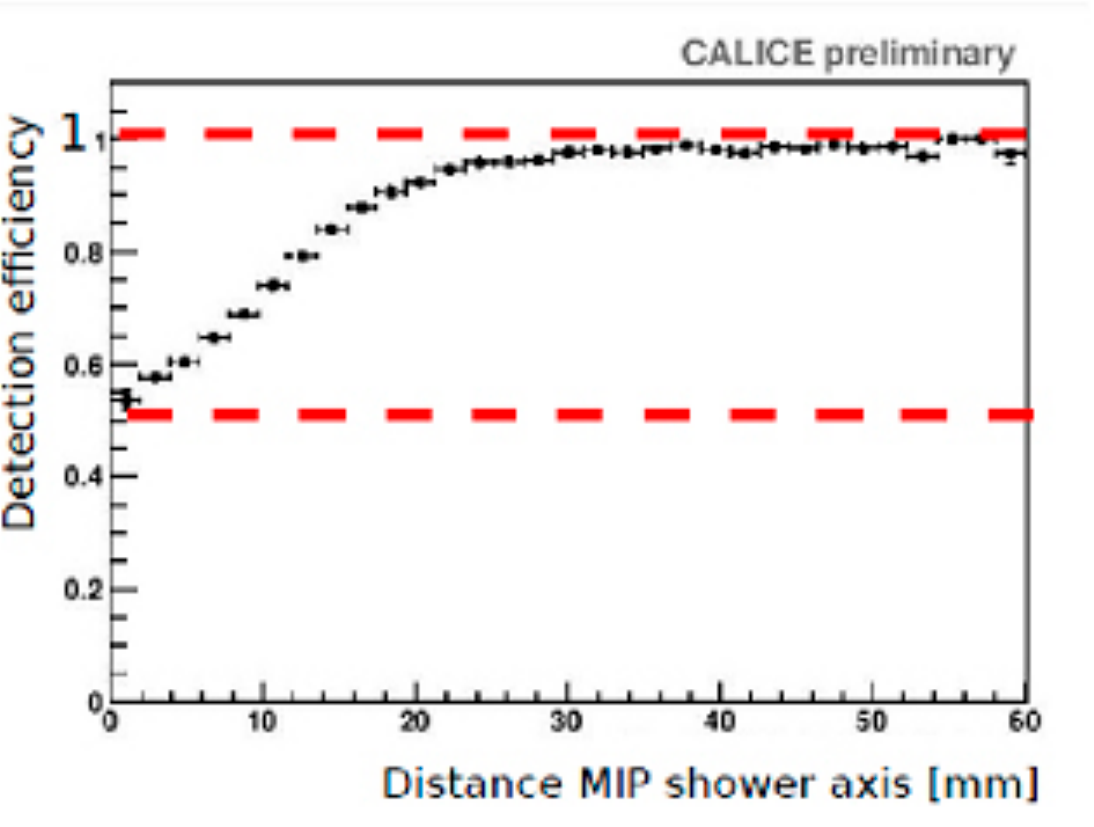}
        \caption{\sl Efficiency of MIP detection as a function of distance from electron shower}
        \label{fig:dist}}
 \end{figure}

The high granularity of the calorimeter allows the tracking of particles as they pass through the detector and the use of imaging processing techniques. For instance, the Hough transform technique has been tested to find tracks in the calorimeter~\cite{fehr2010}. This technique has been applied to reconstruct a muon track near to a 30 GeV electromagnetic shower. The Figure~\ref{fig:dist} shows the reconstruction efficiency as a function of the distance between the track and the shower axis. The efficiency reaches 100\% for distance of 25~mm. This separation efficiency is essential for particle flow algorithm, especially to separate charged and neutral particles which are not detected by trackers.

 \begin{figure}[h]
        {\centering 
                \subfloat{\includegraphics[width=0.45\textwidth]{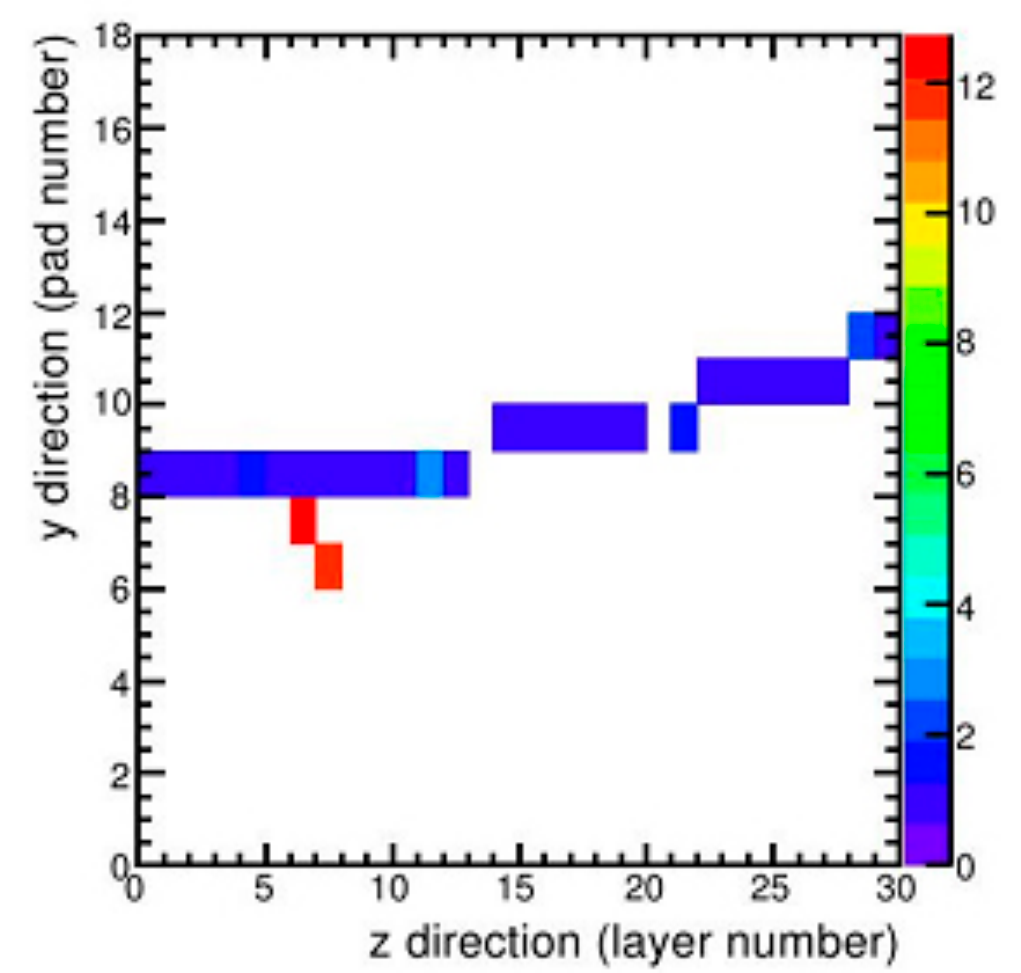}}
                \subfloat{\includegraphics[width=0.45\textwidth]{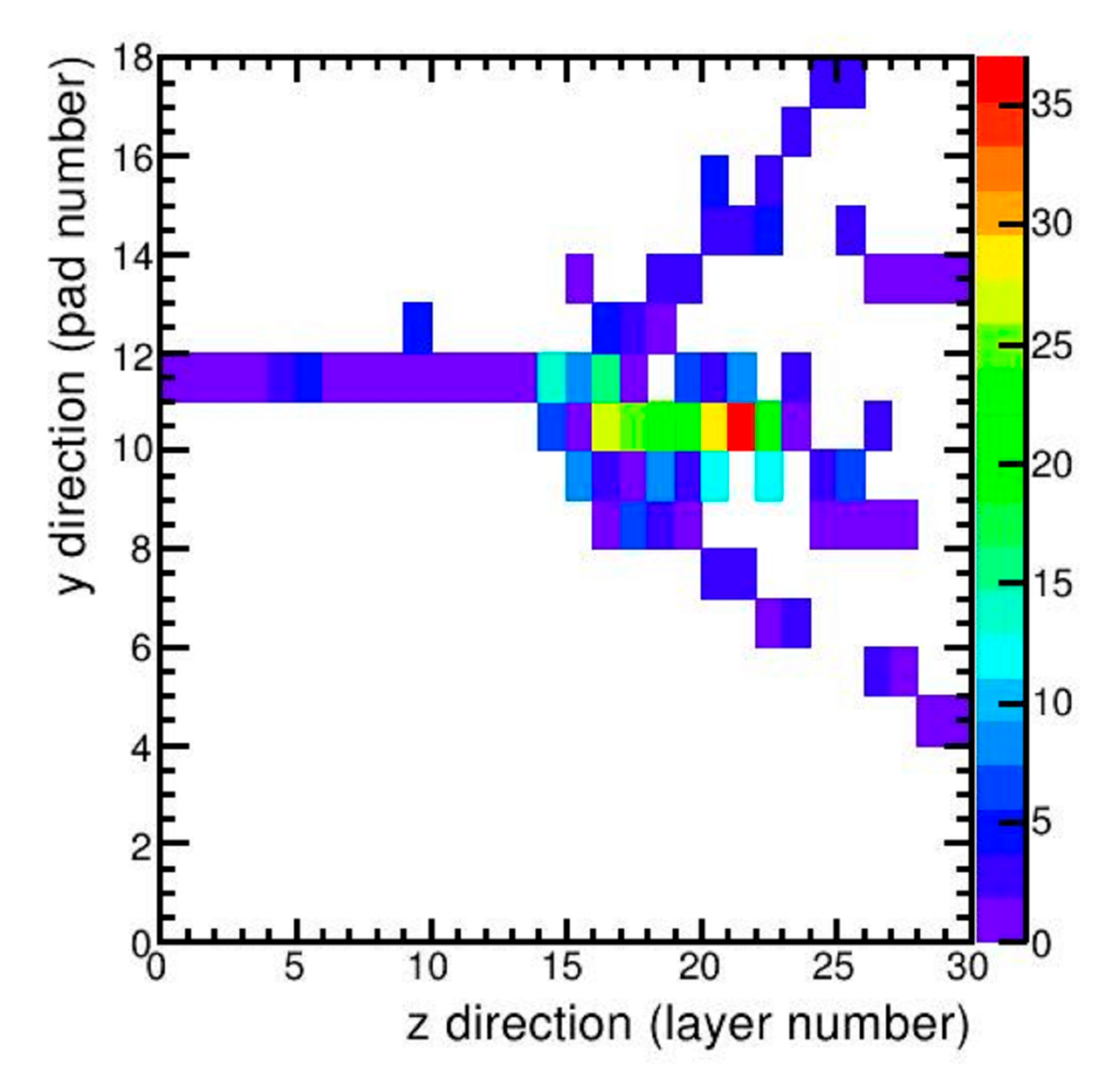}}
        \caption{\sl Examples of hadronic interactions in the ECAL. "FireBall" event on the top and "PointLike" event on the bottom.}
        \label{fig:evt}}
 \end{figure}

 \begin{figure}[h]
        {\centering 
                \subfloat{\includegraphics[width=0.43\textwidth]{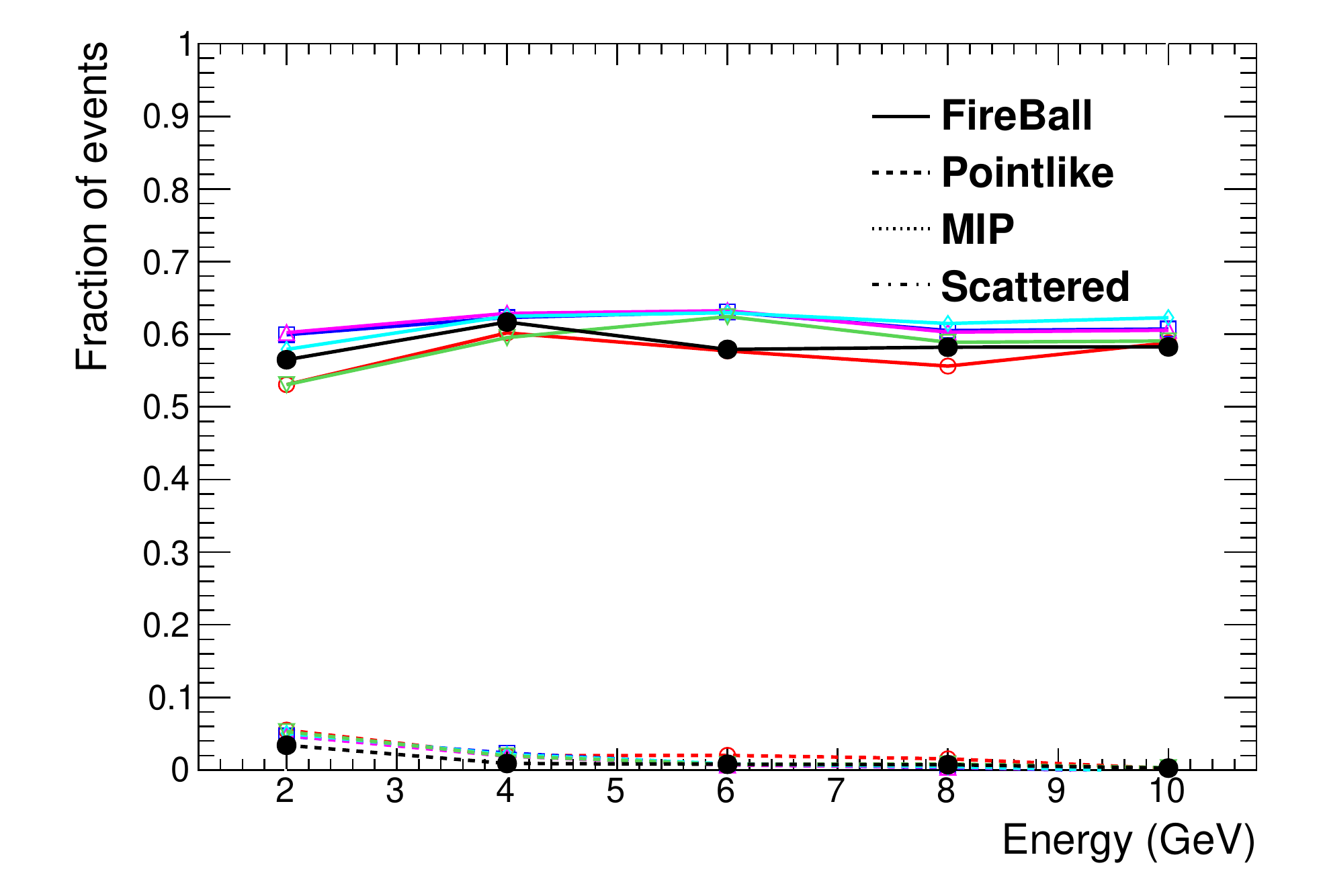}}
                \subfloat{\includegraphics[width=0.43\textwidth]{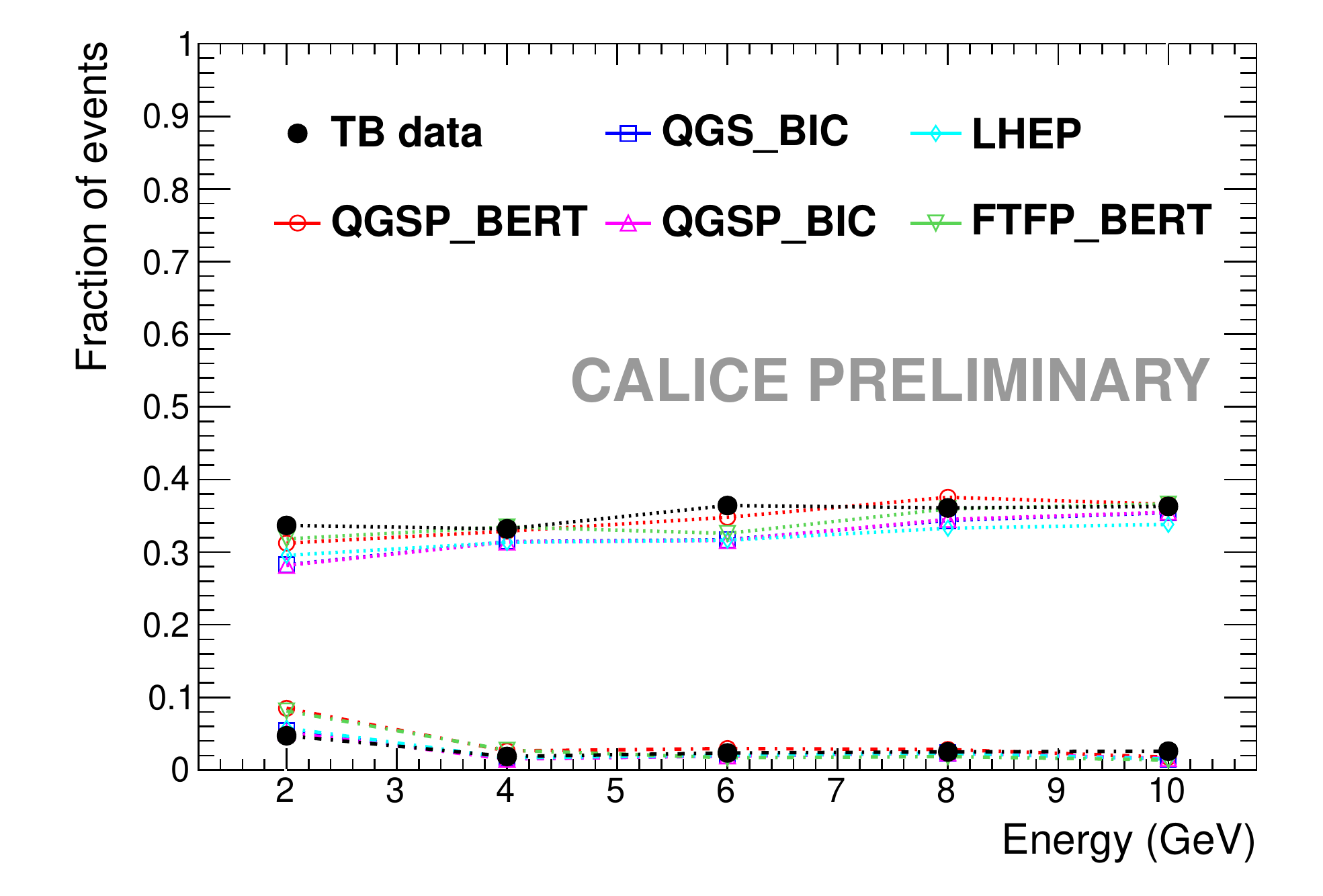}}
        \caption{\sl Frequency of event types and comparison with several physics lists as function of  energies from 2~GeV to 10~GeV. The top figure shows the rates of interactive classes while bottom figure shows the rates of non-interactive classes.}
        \label{fig:frac}}
 \end{figure}

Figure~\ref{fig:evt} illustrates the potential of the high granularity. We distinguish two different and recognizable topologies. Events can be classified as a function of their topologies into 4 classes. MIP and elastic scattering are considered as non interactive events. The two distributions presented in Figure~\ref{fig:evt} show the two types of interacting events called ``Pointlike'' (top) and ``Fireball'' (bottom)~\cite{philippe_thesis}~\cite{doublet2011}.

In Figure~\ref{fig:frac}, we compare the relative frequency of each event class between data and some Geant4 physics lists. The frequency of ``Fireball'' events in the data is about 55\% over the whole energy range. After delta-rays correction, the frequency of ``Pointlike'' events is about 4\% at small energies and tend towards zero at higher energies. The frequency of both classes for inelastic events is well reproduced by all physics lists which confirms that the total inelastic cross sections are well implemented into the physics lists. Cross sections of non-interacting events are also well modeled by Geant4~\cite{philippe_thesis}~\cite{doublet2011}.

\subsection{Transverse profiles}

The shower radius is defined by:
\begin{equation*}
                \displaystyle \mathrm{R}_{\mathrm{E}} = \sqrt{\sigma_{\mathrm{E,x}}^2+\sigma_{\mathrm{E,y}}^2}
\end{equation*}
where:
\begin{equation*}
                \displaystyle \sigma_{\mathrm{E,x}}^2 = \frac{ \displaystyle \sum_{\mathrm{hits}}x_{\mathrm{hit}}^{2}E_{\mathrm{hit}} }{\displaystyle \sum_{\mathrm{hits}}E_{\mathrm{hit}}} - \left( \frac{\displaystyle \sum_{ \mathrm{hits}}x_{\mathrm{hit}}E_{\mathrm{hit}}}{\displaystyle \sum_{ \mathrm{hits}}E_{\mathrm{hit}}} \right)^{2}
\end{equation*}
and the same for y. The transversal profile is an important observable affects the overlap of shower thus the confusion.

For the calculation of the observables, only hits in the interaction layer and all subsequent layers are taken into account. In order to define in the same way a measure of the radius of non interacting events, i.e. interactions where no interaction point could be found, the width R$_{E}$ is calculated by summing over all hits in the ECAL~\cite{philippe_thesis}~\cite{doublet2011}.

\begin{figure}[ht]
\begin{center}
\includegraphics[width=0.47\textwidth]{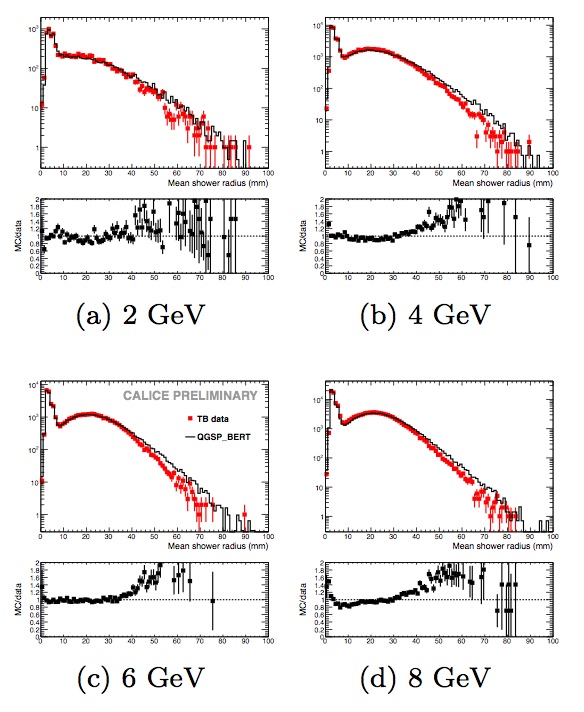}
\includegraphics[width=0.49\textwidth]{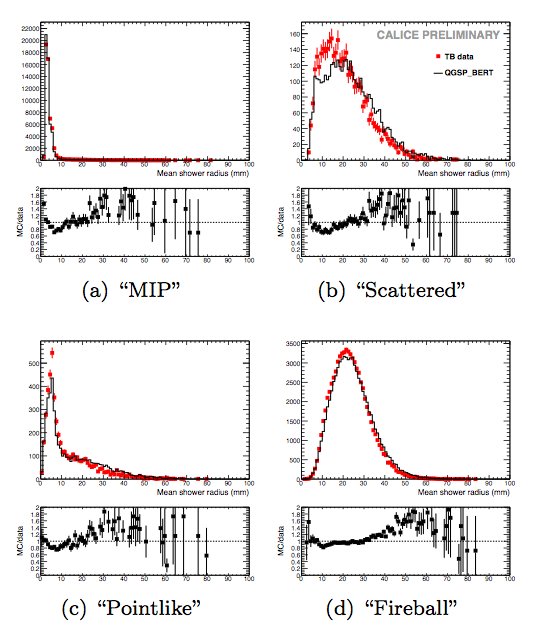}
\caption{\sl \underline{Left:} Lateral distributions of the hadronic shower energy at energies from 2 GeV to 10 GeV. For each distribution: the top view features the comparison between test beam data in red and the simulation with QGSP\_BERT in black. The bottom view shows the ratio of test beam data over simulation. \underline{Right:} Lateral distributions of the hadronic shower for each event classes at 8 GeV. For each distribution: the top view features the comparison between test beam data in red and the simulation with QGSP\_BERT in black. The bottom view shows the ratio of test beam data over simulation.
}
\label{fig:traprof}
\end{center}
\end{figure}

The left part of Figure~\ref{fig:traprof} shows the comparison of the transverse profile between data and simulation at energies from 2~GeV up to 10~GeV. In this plot we use the QGSP\_BERT physic list. We distinguish 2 maxima at 5~mm and 20~mm. Up to 6~GeV, data are well reproduced by the simulation except for radius larger than 35~mm. At 8~GeV the simulation overestimate the data in the transition region between the two peaks. The situation is comparable at 10\,GeV

In the right part of Figure~\ref{fig:traprof}, the contribution of each event type is separated for 8 GeV events. We see that the first maximum in Figure~\ref{fig:traprof} left is created by MIP events and the second one by  ``Fireball'' events. The maxima of ``Pointlike'' and ''Scattered'' events are located in the transition region. The four event types show the same disagreement~\cite{philippe_thesis}~\cite{doublet2011}.

\subsection{Longitudinal profiles}

 \begin{figure}[h]
        {\centering 
                \subfloat{\includegraphics[width=0.4\textwidth]{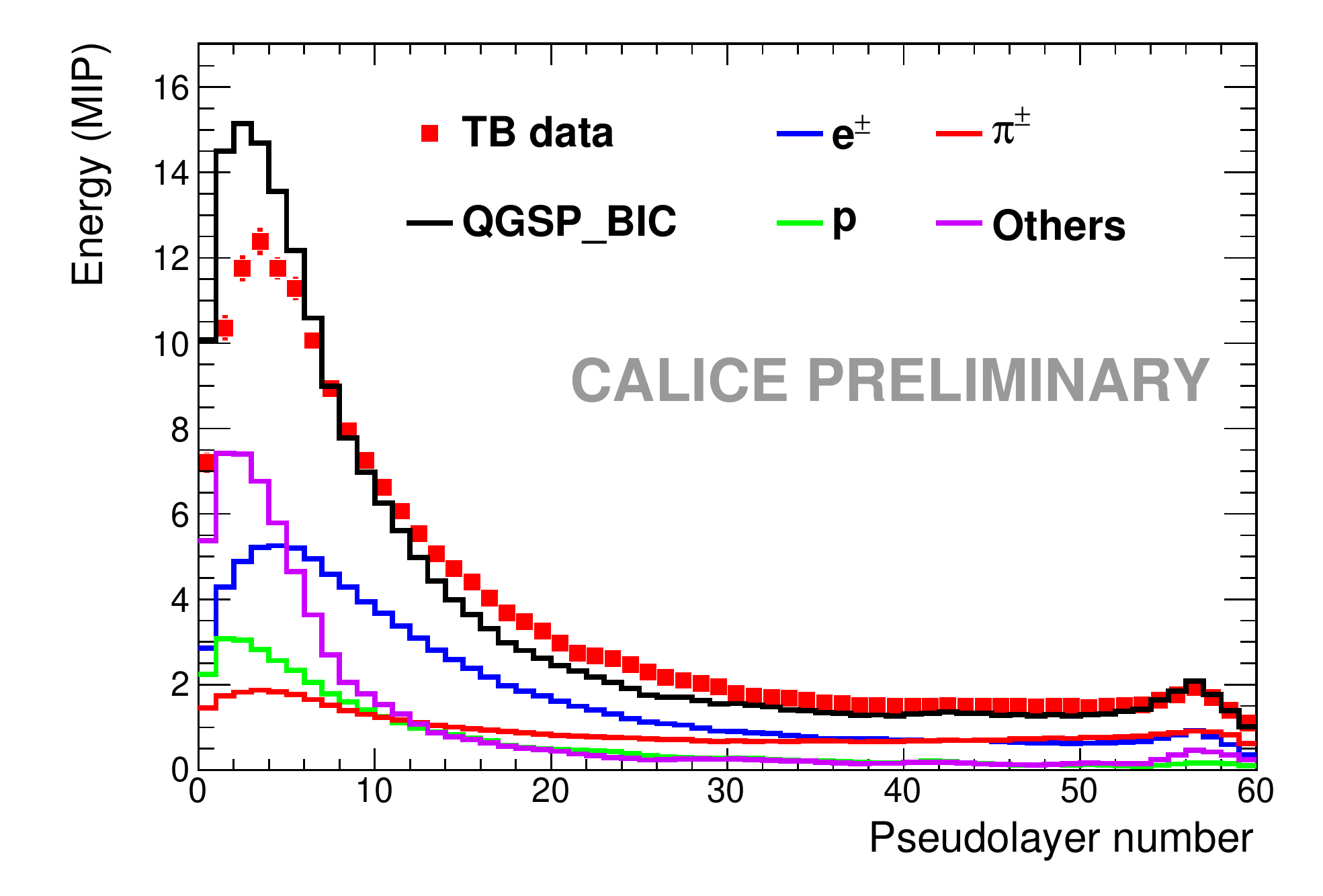}}
                \subfloat{\includegraphics[width=0.4\textwidth]{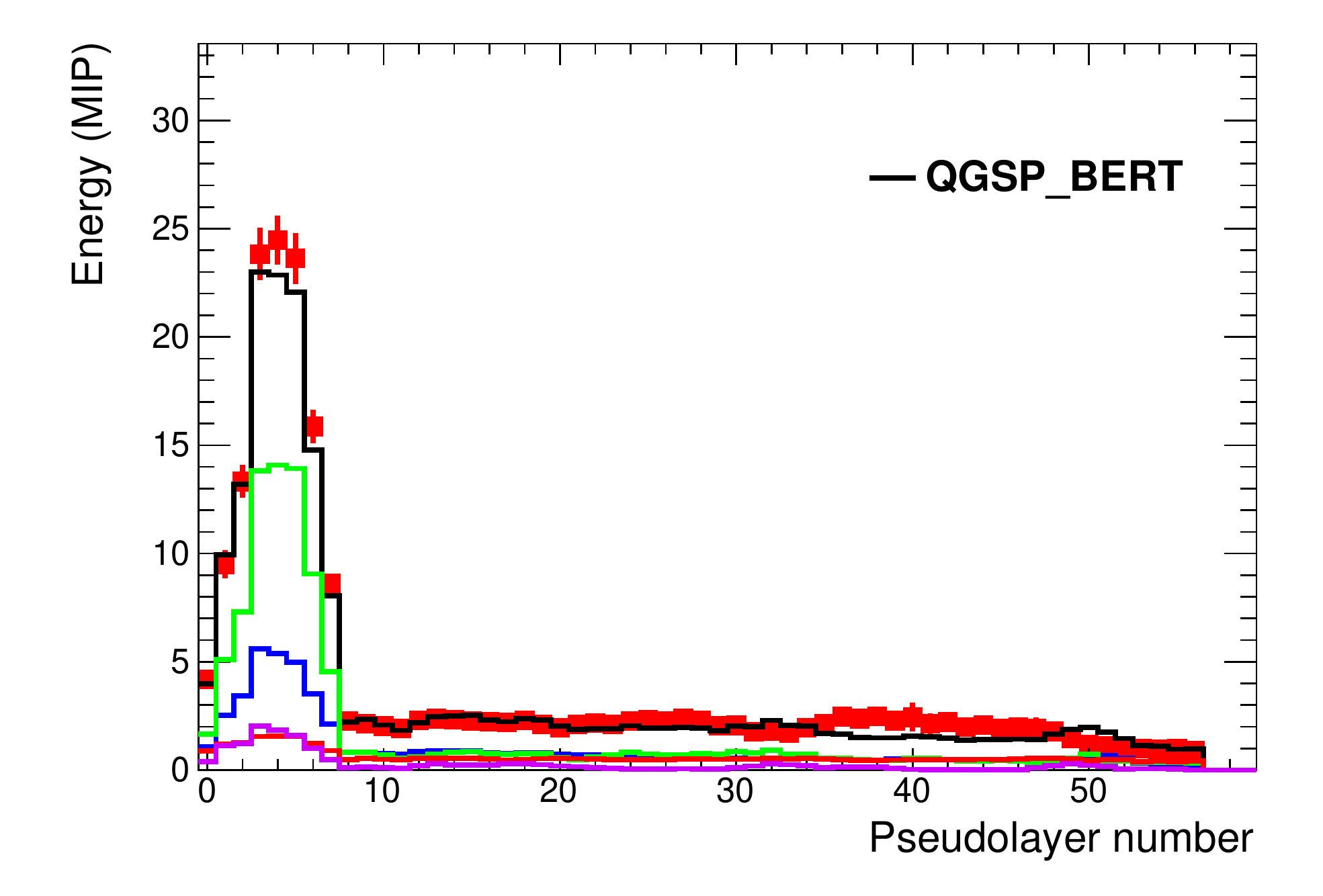}}
        \caption{\sl Total longitudinal profiles for the two interacting event types: comparison between test beam
 data in red and simulation in black at 2~GeV. For simulated events, we separated the contributions of the secondary particles. The blue lines are the contributions from electrons and positrons. The green lines are the contributions from protons. The red lines are the contributions of pions. Other particles are in violet. The top figure shows the comparison between test beam data and the QGSP\_BIC physic list for "Fireball" events while the bottom figure shows the comparison between test beam data and the QGSP\_BERT physic list for "Pointlike" events.}
        \label{fig:prof}}
 \end{figure}

The longitudinal profile is given as a function of pseudolayers in order to account for the different sampling fractions in the ECAL. In the first module, pseudolayers are equivalent to real layers. In the second module, each layer is subdivided in two pseudolayers and in the third module, layers are subdivided into three pseudolayers. The energy is linearly interpolated between the pseudolayers. In case of interacting events, the longitudinal shower profile starts from the interaction layer. In case of non-interacting events, the longitudinal profile is calculated from the first detector layer~\cite{philippe_thesis}~\cite{doublet2011}.

In Figure~\ref{fig:prof}, the test beam data are shown in red and the simulation in black. For simulated events, we separated the contributions of the secondary particles. The blue lines are the contributions from electrons and positrons. The green lines are the contributions from protons. The red lines are the contributions of pions. Other particles are in purple. For "Fireball" events, none of the models give a satisfactory description of the data. For instance, the top distribution shows significant disagreement between data and the QGSP\_BIC model. In the bottom part of Figure~\ref{fig:prof}, we present longitudinal profile of "Pointlike" events and the comparison with QGSP\_BERT model. This class of events is better reproduce by the Bertini model~\cite{philippe_thesis}~\cite{doublet2011}.

\section{Conclusion and outlook}

This study investigates in depth the interactions of pions in an energy range between 2 and 10\,GeV. It demonstrates the outstanding potential of the CALICE silicon tungsten electromagnetic calorimeter to resolve details of hadronic cascades. The data obtained in test beams are compared with different physics lists provided by Geant4. Visible differences between models appear at this level of detail. The study on shower decomposition is the first step toward the use of sophisticated algorithms which may be developed in collaboration with experts from applied mathematics. To succeed the physics prototype, a technological prototype is in development which will feature a four times higher granularity.~\cite{techprot}

\section{Acknowledgments}
Thanks to Thibault Frisson who has written the major part of these proceedings. The text is based on the PhD thesis of Philippe Doublet.
%This work is funded by the European Union in the 6th framework program "Structuring the European Research Area", the 7th framework program "Capacities" as well as by the CNRS/IN2P3 budget for "Quarks and Leptons", the French ANR program and the Japanese JSPS program

\bibliographystyle{unsrt_tibo}

\begin{footnotesize}
\bibliography{biblio}
\end{footnotesize}

\end{document}